\documentclass{article}
\usepackage[T1]{fontenc}
\usepackage{wrapfig}
\usepackage[portuges,english]{babel}
\usepackage{amssymb,latexsym,amsmath,color,mathrsfs,ifsym,graphics,stmaryrd}
\usepackage{amsthm, amsfonts, mathrsfs}
\usepackage[colorlinks,linkcolor=blue,urlcolor=blue,citecolor=black,
plainpages=false,pdfpagelabels,breaklinks]{hyperref}
\usepackage[all]{xy}
\usepackage{caption}
\usepackage{graphicx}
\usepackage[margin=2.2 cm]{geometry}

\DeclareMathOperator{\EA}{EA}

\begin{document}

\title{{\sc On the Relative Nature of Quantum Individuals}}

\author{{\sc Christian de Ronde}$^{1,2,3}$, {\sc Raimundo Fern\'andez Mouj\'an}$^{2,6}$, {\sc C\'esar Massri}$^{4,5}$}

\date{}

\bibliographystyle{plain}
\maketitle

\begin{center}
\begin{small}
1. Philosophy Institute Dr. A. Korn, University of Buenos Aires - CONICET\\
2. Center Leo Apostel for Interdisciplinary Studies\\Foundations of the Exact Sciences - Vrije Universiteit Brussel\\
3. Institute of Engineering - National University Arturo Jauretche\\
4. Institute of Mathematical Investigations Luis A. Santal\'o, UBA - CONICET\\
5. University CAECE\\
6. Philosophy Institute, Diego Portales University, Santiago de Chile
\end{small}
\end{center}

\begin{abstract}
\noindent In this work we argue against the interpretation that underlies the ``Standard'' account of Quantum Mechanics (SQM) that was established during the 1930s by Niels Bohr and Paul Dirac. Ever since, following this orthodox narrative, physicists have dogmatically proclaimed --quite regardless of the deep contradictions and problems-- that the the theory of quanta describes a microscopic realm composed of elementary particles (such as electrons, protons and neutrons) which underly our macroscopic world composed of tables, chairs and dogs. After critically addressing this atomist dogma still present today in contemporary (quantum) physics and philosophy, we present a new understanding of {\it quantum individuals} defined as the minimum set of relations within a specific {\it degree of complexity} capable to account for all relations within that same degree. In this case, quantum individuality is not conceived in absolute terms but --instead-- as an {\it objectively relative} concept which even though depends of the choice of bases and factorizations remain nonetheless part of the same invariant representation.  
\end{abstract}
\begin{small}

\end{small}

\newtheorem{theo}{Theorem}[section]
\newtheorem{definition}[theo]{Definition}
\newtheorem{lem}[theo]{Lemma}
\newtheorem{met}[theo]{Method}
\newtheorem{prop}[theo]{Proposition}
\newtheorem{coro}[theo]{Corollary}
\newtheorem{exam}[theo]{Example}
\newtheorem{rema}[theo]{Remark}{\hspace*{4mm}}
\newtheorem{example}[theo]{Example}
\newcommand{\ninv}{\mathord{\sim}} 
\newtheorem{postulate}[theo]{Postulate}
\newcommand{\Proof}{\textit{Proof:} \,}
\newcommand{\cqd}{\hfill{\rule{.70ex}{2ex}} \medskip}

\bigskip

\bigskip

\bigskip

\bigskip

\bigskip

\section{The Microscopic Realm of ``Standard'' Quantum Mechanics}

As it appears to us, the discussion on individuality, in the context of the debates on quantum theory, has been determined, almost from the beginning, by a basic presupposition, by a fundamental representation, which, partly due to its familiarity, and partly due to the work of Niels Bohr, managed to permeate the general understanding of quantum physics. We are talking, of course, about an atomist representation of the world, about an atomist metaphysics which, although undoubtedly suitable for Newtonian mechanics, has also been considered a priori true when reflecting on Quantum Mechanics (QM). In this generally accepted view, QM talks about the microscopic realm, which is populated by quantum particles\footnote{The idea that the fundamental aspect of the quantum theory signifies its application to the microscopic is in fact merely one possible interpretation --and not the only one-- of that fundamental character. And it is undoubtely an interpretation which is dependent on the atomist representation.}. And the problem, which then becomes a factory of other problems, is that such representation, which entails the idea of small substantial bodies in space that in their sum constitute reality, is incompatible with the formalism and the experience inherent to the theory of quanta. Indeed, since the establishment of the Standard formulation of QM (SQM) during the early 1930s there has existed, within the physics community, a general consensus regarding the claim that the theory of quanta describes a microscopic realm composed of elementary quantum particles --such as electrons, protons, neutrons and the like particles--. Ever since, the reference to these microscopic entities has grounded the analysis of the theory  becoming a cornerstone of the different lines of research in quantum physics. At the same time, however, something rather curious happened with this atomist representation in the context of QM: mainstream physicists would also claim that ``nobody understands quantum mechanics'' --a statement made popular by Richard Feynamnn during the 1960s \cite{Feynman67}. Even today, quantum physicists, at least when pushed hard enough, will end up either recognizing that they are truly incapable of explaining what {\it is} a quantum particle or accepting that their discourse about particles is ``just a way of talking'' (see for a detailed discussion \cite{deRondeFM21, Wolchover20}). This  paradoxical characteristic of the discourse regarding QM (the confidence in a representation nobody understands) can in fact be traced all the way back to the origins of the theory, specifically, to Bohr's work. 
Trying to make the atomist representation of the world compatible with the new discoveries of the physics of the first decades of the 20th century, Bohr proposed in 1913 a new atomic model which combined a familiar image (elementary particles orbiting around a nucleus like planets around the sun) with a new set of strange {\it ad-hoc} rules, such as mysterious ``jumps'' performed by electrons between orbits. Although no real proof or justification was presented for the existence of these ``jumping'' particles, the physics community was immediately captivated by Bohr's images. But not everyone: as it is well known, Albert Einstein and Erwin Schrödinger were rather suspicious of these unjustified aspects of Bohr's interpretation. The curious way in which made up fictions such as ``quantum particles'' and ``quantum jumps'' were applied by Bohr --quite regardless of any theoretical representation or experimental evidence-- was exposed by Werner Heisenberg in his autobiography. The German physicist was a direct witness of a meeting between Bohr and Schr\"odinger, which took place in Copenhagen in 1926, in order to discuss the existence of ``quantum jumps'' within Schr\"odinger's new wave formulation. As Heisenberg would recall, even though the many arguments that Schr\"odinger \cite[p. 73]{Heis71} had produced during the debate had allowed him to rationally conclude that ``the whole idea of quantum jumps is sheer fantasy'', the Danish illusionist, with a single move of his ``magic wand''\footnote{A term applied by Arnold Sommerfeld in order to characterize Bohr's methodology \cite{BokulichBokulich20}.}, would invert the burden of proof turning things completely upside-down: 
\begin{quotation}
\noindent {\small ``What you say is absolutely correct. But it does not prove that there are no quantum jumps. It only proves that we cannot imagine them, that the representational concepts with which we describe events in daily life and experiments in classical physics are inadequate when it comes to describing quantum jumps. Nor should we be surprised to find it so, seeing that the processes involved are not the objects of direct experience.'' \cite[p. 74]{Heis71}} 
\end{quotation}
One of the main cornerstones of Bohr’s program is the combination of, on the one hand, a narrative according to which QM talks about a ``microscopic'' physical realm constituted by ``elementary particles'', and, on the other hand, the idea that these elementary particles escape theoretical representation, that it is impossible to conceptually apprehend this postulated microscopic realm. In this manner, Bohr astutely accomplished the imposition of his atomist narrative, and, by convincing physicists of the irrepresentability of quantum particles, he was able at the same time to avoid any critical analysis of his atomist discourse. He imposed his view and, at the same time, he denied the possibility of the critical revision of that view. A view that was generally accepted but, at the same time, one that nobody could really explain or justify.
The last quote of Heisenberg's book is a great example of Bohr’s tactics, as just explained: he postulated the existence of ``quantum jumps'', and then rejected Schrödinger’s critique of these postulated phenomena by referring to their irrepresentability. After the meeting, confessing his impotency, Schrödinger would write to his friend Wilhelm Wien: 
\begin{quotation}
\noindent {\small ``Bohr's [...] approach to atomic problems [...] is really remarkable. He is completely convinced that any understanding in the usual sense of the word is impossible. Therefore the conversation is almost immediately driven into philosophical questions, and soon you no longer know whether you really take the position he is attacking, or whether you really must attack the position he is defending.'' \cite[p. 228]{Moore89}} 
\end{quotation}   
In any case, this has helped to uncritically retain the fundamentally unjustified claim that QM talks about elementary particles such as electrons, protons and neutrons. This atomist presupposition is still today one of the main obstacles for the understanding of the theory of quanta. One may take this idea as an exaggeration, claiming --as we already mentioned above-- that quantum particles are ``just a way of talking''. But, in fact, the problem is deeper, as the atomist supposition implies, wether we like it or not, a series of deductions, and methodological and operational steps, that determine right from the start the understanding of the formalism and of observations. As Faraday explained long ago: ``the word \emph{atom}, which can never be used without involving much that is purely hypothetical, is often intended to be used to express a simple fact; but good as the intention is, I have not yet found a mind that did habitually separate it from its accompanying temptations'' \cite[p. 220]{Laudan81}. Schrödinger rephrases this idea for the quantum case: ``We have taken over from previous theory the idea of a particle and all the technical language concerning it. This idea is inadequate. It constantly drives our mind to ask information which has obviously no significance'' \cite[p. 188]{Schr50}. It is not difficult to understand that if one dogmatically applies from the beginning, as a starting point, consciously or not, a series of categorical and formal principles --such as  set-theory, or those of particle metaphysics, e.g., separability, individuality, locality, etc.-- to a mathematical formalism that was never meant to be understood under the constraints of such representation, the result of this methodology will lead only to paradoxes and dead ends.

Of course, by now, contemporary physicists which have been trained for many decades in an instrumentalist fashion do not even consider the consistent and coherent description of quantum particles as a meaningful scientific problem --although they still presuppose an atomist discourse with all the deductions and methodological steps which it implies--. After all, they argue, SQM is already capable, when considered as an algorithmic ``recipe'', to predict the measurement outcomes that are required in order to produce new technology (see \cite[pp. 2-3]{Maudlin19}). And thus, when students make questions about the reference of the theory they are kindly advised to ``Shut up! And calculate!'' 
In this context, it is philosophers of QM which, since the constitution of the field during the 1980s, have taken this problem to be part of their own specific field of expertise. Thus, it is only within the philosophical arena that researchers have attempted to account more explicitly for the microscopic realm supposedly described by QM through the addition of different ``interpretations'' --some of which go even beyond this microscopic discourse. But even though these narratives add to the standard formulation new complex layers to the ground zero narrative already applied by physicists, they do not attempt to replace or confront the main aspects of this ``original atomist interpretation'' which has also founded, as it is well known, some of the most important mainstream lines of research within contemporary experimental and theoretical physics such as the {\it Standard Model of Particles} or {\it String Theory}. It should be clear that regardless of the interpretational debate --which has problems of its own--, no one, physicist or philosopher, fundamentally disputes the orthodox inconsistent discourse of SQM which claims to {\it know} and {\it not know} that the theory talks about microscopic particles. As a mater of fact, physicists actively apply this atomist picture in many different models discussed within the theory, such as entanglement, decoherence, quantum teleportation, quantum cryptography, etc. And philosophers of QM do exactly the same when addressing the work of physicists or their own interpretational problems (e.g., the quantum to classical limit, the measurement problem or even the characterization of quantum particles themselves in negative terms, talking about their non-separability, non-locality, non-individuality, etc.). Thus, what needs to be clear is that even though philosophers do explicitly ``add'' new supplementary narratives that talk about parallel worlds, parallel minds, the existence of quantum fields, potentialities, propensities, flashes, etc., they do not engage in a fundamental critical analysis of the atomist foundation that was established by physicists almost a century ago and continues even today to provide the starting point of mainstream contemporary scientific research. 
In any case, this has resulted in a universal presupposition that, consciously or not, permeates all approaches to QM: a representation of individual separated substances in space which in their sum constitute reality. Although one can find some exotic interpretations which add complex layers to this main representation, the truth is that this fundamental representation, with its natural ``temptations'', continues to function on the fundamental level. And the problem is that it is exactly that fundamental representation that is incompatible with the quantum formalism and with experience.

A both perfect and paradoxical symptomatic expression of the consequences of applying a priori the atomist representation to QM can be found in a current discussion in the philosophy of QM: the one concerning the supposed non-individuality of quantum particles (see for a detailed analysis \cite{Arenhart17, Arenhart23, KrauseArenhart18}). By assuming that QM speaks of particles, the immediate problem arises that these particles, that one should be able to distinguish, cannot be distinguished; they appear as indistinguishable. Their sameness, their identity, is completely lost. This problem --let it be clarified-- arises only because particles are assumed beforehand. Now, instead of taking this fact as proof that the theory speaks of entities that are entirely of a different nature than particles, a curious question is raised: are quantum particles individuals or non-individuals? Instead of undertaking the task of providing concepts for a type of originally quantum individual, different from the particle, there is a progressive erasure of the referent,  which, of course, nobody really knows what it means, because all that is being done is to assume a type of entity while simultaneously denying one of its essential characteristics. A type of entity is assumed and at the same time emptied of what defines it. Along this path, it seems that little more than an empty concept remains in our hands. But the truth is that \emph{there are} individuals, and there is no need to strip them of characteristics; it's just that these individuals are entirely of a different nature and must be thought of with other categories. It is not a matter of starting from the atomist assumption to gradually empty its entities of their own characteristics, but rather of thinking --setting aside the assumed atomist metaphysics-- about what kind of individuality is capable of being consistently related to the mathematical formalism and, at the same time, of producing a consistent and coherent understanding of quantum phenomena.

\section{The Atomist Myth within Quantum Physics}

The substantialist atomist interpretation is found everywhere within the orthodox mainstream research in quantum physics. This, regardless of the fact that those same physicists might also claim --embracing an inconsistent discourse-- that they do not actually know --apart from stressing its ``weirdness''-- what the theory really talks about. Let us provide some examples of the way in which contemporary physicists do include explicitly within their research the reference to particles in order to ``explain'' what they actually investigate. 

One obvious example is quantum decoherence, a process of environmental ``loss of coherence'' that supposedly turns quantum superpositions into classical systems and  where the reference to particles appears explicitly right from the start. As explained by the Argentine physicist Juan Pablo Paz: 
\begin{quotation} 
\noindent {\small ``At the atomic level, electrons and protons are blurred entities that cannot be described as point-like particles following trajectories. But macroscopic objects have well defined properties: they are either here or there, and not everywhere. So how does the classical world arise from the quantum?

The consensus today is that classical behaviour is an emergent property of quantum systems, induced by their interaction with the environment. This interaction, a fact of life for complex macroscopic objects, is responsible for the process of decoherence. \cite[p. 869]{Paz01}} 
\end{quotation} 
The philosopher of physics, Guido Bacciagaluppi also refers to particles:   
\begin{quotation} 
\noindent {\small ``So, for example, there could be sufficiently many stray particles that scatter off the electron. The phase relation between the two components of the wave function, which is responsible for interference, is now well-defined only at the level of the larger system composed of electron and stray particles, and can produce interference only in a suitable experiment including the larger system. Such a phenomenon of suppression of interference is what is called decoherence.'' \cite{Bacc12}} 
\end{quotation}
However, as Bacciagaluppi also remarks, when posing the question: ``can we derive from quantum mechanics the behaviour that is characteristic of [...] `classical' systems?'' The answer is clear: ``such a derivation appears impossible. To put it crudely: if everything is in interaction with everything else, everything is generically entangled with everything else, and that is a worse problem than measuring apparatuses being entangled with measured systems.'' As it is well known in the philosophical literature, the {\it environment} introduced by decoherence, supposedly conformed by many elementary particles, detaches itself right from the start from the quantum theoretical representation in terms of quantum superpositions and is described --instead-- in purely classical terms, as a {\it continuous bath}. Thus, what needs to be ``explained'', namely, the path from a {\it discrete} quantum representation to a {\it continuous} classical description of physical reality is dogmatically imposed. In fact, during the 1990s, within the philosophical literature, and quite regardless of the many efforts to argue in favor of the existence of decoherence, the program was exposed as a complete failure. Unfortunately, the acceptance of this failure in the mainstream literature would be soon renamed as a new ``solution for all practical purposes'', in short, ``FAPP''. Regardless of this debate, what is clear is that the reference to particles is essential within these models. 

The reference to the microscopic realm appears as well within the new field of quantum information processing. There, even in the case of teleportation the reference to quantum particles is also explicit:
\begin{quotation}
\noindent {\small``Roughly, the impossibility [of unconditionally secure two-party bit commitment based solely on the principles of QM] arises because at any step in the protocol where either Alice or Bob is required to make a determinate choice (perform a measurement on a particle in the quantum channel, choose randomly and perhaps conditionally between a set of alternative actions to be implemented on the particle in the quantum channel, etc.), the choice can delayed by entangling one or more `ancilla' (helper) particles with the channel particle in an appropriate way. By suitable operations on the ancillas, the channel particle can be `steered' so that this cheating strategy is undetectable.'' \cite{Bub97} } \end{quotation}

But one of the best examples --which even provides the bases for quantum information-- can be observed when addressing one of the most important contemporary notions of the theory of quanta. The history of quantum entanglement can be seen as a clear symptom of the current situation within mainstream physical research. As it is well known, the notion of entanglement had a critical origin, it was developed, by Einstein as well as by Schrödinger, as a sort of \emph{reductio ad absurdum}, as a proof of the inconsistencies of SQM. As explained by Alisa Bokulich and Gregg Jaeger  \cite[p. xiv]{BokulichJaeger10}:  ``[in the EPR paper] the possibility of such a phenomenon [of entanglement]  in QM was taken to be a {\it reductio ad absurdum} showing that there is a fundamental flaw with the theory.'' However, since the 1990s, disregarding completely the work of Einstein and Schr\"odinger, entanglement has simply been accepted, without engaging in a fundamental critical task, without really taking into considerations the incompatibility between this phenomena and the standard representation of the theory. The absurd was simply accepted as meaningful, the problem was, without further analysis, taken to be a solution. And entanglement has been ever since understood, using the language of SQM, in terms of ``the  non-separable nature of quantum particles represented by pure states''. Indeed, today, almost every paper about quantum entanglement begins with the unavoidable reference to ``quantum particles'', something which is presented as an unquestionable element of the theory. Only in some cases, an attempt to avoid the reference to particles is made through the euphemistic reference to ``quantum  systems'', or simply to ``quantum objects''. But a simple change of words, without a comprehensive critical analysis, is not enough, as those deduction, methodological and operational steps, originated in the atomist presupposition, continue to function. In any case, the reference to particles is present --explicitly or implicitly-- in the introduction to almost every published paper about quantum entanglement. Just to give a few examples coming from some of the most prestigious researchers in the field:
\begin{itemize}
\item Abner Shimony \cite{Shimony95}: ``A quantum state of a many-particle system may be `entangled' in the sense of not being a product of single-particle states.'' 
\item William K. Wootters \cite{Wootters98}: ``Quantum mechanical objects can exhibit correlations with one another that are fundamentally at odds with the paradigm of classical physics; one says that the objects are `entangled'.''
\item Jian-Wei Pan, Dik Bouwmeester, Harald Weinfurter, and Anton Zeilinger \cite{Zeilinger98}: ``entanglement has been realized either by having the two entangled particles emerge from a common source, or by having two particles interact with each other. Yet, an alternative possibility to obtain entanglement is to make use of a projection of the state of two particles onto an entangled state.''
\item Ryszard, Pawe, Micha and Karol Horodecki \cite{Horodecki09}: ``[Entanglement is an] holistic property of compound quantum systems, which involves nonclassical correlations between subsystems.''
\item Vlatko Vedral \cite{Vedral14}: ``entanglement can exist in many-body systems (with arbitrarily large numbers of particles).'' 
\item Thomas, R.A., Parniak, M., Ostfeldt, C. et al. \cite{Thomas21}: ``Entanglement is an essential property of multipartite quantum systems, characterized by the inseparability of quantum states of objects regardless of their spatial separation.''
\item Richard Cleve and Harry Buhrman \cite{CleveBuhrman97}: ``If a set of entangled particles are individually measured, the resulting outcomes can exhibit `nonlocal' effects. These are effects that, from the perspective of `classical' physics, cannot occur unless `instantaneous communications' occur among the particles, which convey information about each particle's measurement to the other particles.''
\item  Davide Castelvecchi and Elizabeth Gibney \cite{Nature22}: ``Because of the effects of quantum entanglement, measuring the property of one particle in an entangled pair immediately affects the results of measurements on the other. It is what enables quantum computers to function: these machines, which seek to harness quantum particles' ability to exist in more than one state at once, carry out calculations that would be impossible on a conventional computer.'' 
\end{itemize}
It is interesting to note something that becomes clear from several of the different definitions of entanglement mentioned above: it is a notion that is only addressed in terms of its inability to satisfy the criteria of an atomist representation more typical of classical physics. That is, in reality, there is no positive definition given of what entanglement {\it is}, but only a negative one, which states what entanglement {\it is not}, what it fails to be. But let us see one specific reason that explains why the orthodox atomist approaches to entanglement fail. 

The mythical reference to `particles' which grounds the interpretation of {\it factorizability} --understood as the separation of systems into sub-systems-- is, from a conceptual and formal perspective of analysis, essentially inconsistent \cite{deRondeFMMassri24a, deRondeMassri23, Earman15}. The notion of separability is grounded on the modern metaphysical atomist representation provided by classical physics according to which physical reality is composed of independent separated individual entities which exist within space and time. According to this supposedly ``commonsensical'' picture, a system can be understood in terms of its parts and the knowledge of these parts implies the knowledge of the whole. This is of course a direct consequence of the underlying Boolean logic that is a prerequisite to define the notion of entity understood classically. Indeed, as a consequence, the propositions derived from classical mechanics can be arranged in a Boolean lattice (see for a detailed discussion \cite{deRondeFreytesDomenech18}). According to classical logic, and following set theory, the {\it sum} or {\it union} of the elements of a system imply its complete characterization as a whole. It is easy to picture all of this in terms of sets and the logical relations that we learned at school when we were little: 

\begin{center}
\includegraphics[scale=.3]{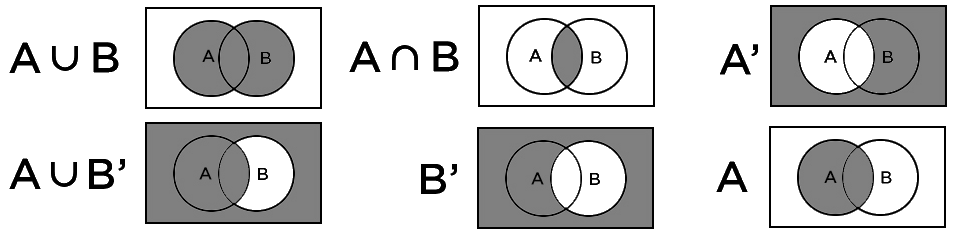}
\captionof{figure}{Union, intersection and complement in Boolean logic.}
\end{center}

\noindent However, as it is also well known since the famous paper by Birkhoof and von Neumann \cite{BvN36}, the underlying logic of QM is not Boolean, it is not {\it distributive}. Thus, the basic classical way of reasoning about systems becomes precluded right from the start. This is an obvious consequence of the fact that vectorial spaces do not relate between each other following the same rules as the elements of a set through {\it union} and {\it conjunction}. In the quantum case the equivalent to the {\it union} of two vectors is not the {\it sum} of the individual vectors considered as lines, but instead what they are capable to {\it generate} in terms of subspaces. Thus, in the particular case when we consider the {\it sum} of two vectors what we obtain is the whole plane.

\begin{center}
\begin{tabular}{ccc}
\includegraphics[scale=.5]{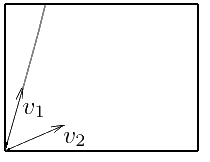} &
\includegraphics[scale=.5]{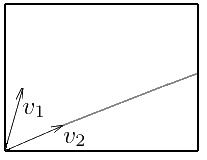} &
\includegraphics[scale=.5]{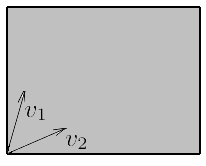} \\
$\langle v_1\rangle$  &
$\langle v_2\rangle$ &
$\langle v_1\rangle\oplus\langle v_2\rangle$
\end{tabular}
\end{center}
As it could have been easily foreseen, the artificial {\it ad hoc} introduction of a set of logical relations completely alien and even incompatible with the mathematical formalism of the theory could only lead to confusions, contradictions and pseudo-problems. Sadly enough, this is exactly what happened with the introduction of the notion of separability in the context of QM. As we just explained, the \emph{union} of two vectors was inadequately understood as a \emph{sum} when, in fact, it is a \emph{generation}. Analogously, the \emph{projection} of a subspace was incorrectly interpreted as a {\it separation} of the whole set and the choice of a subset of elements, when, as a matter of fact, its correct interpretation is that of {\it shadow} \cite{deRondeMassri23}. In figure 2 we can see that while the shadow of $|\Psi\rangle$ in the x-axis is the vector $|x\rangle$, the shadow of $|\Psi\rangle$ in the y-axis is the vector $|y\rangle$.

\begin{center}
\includegraphics[scale=.3]{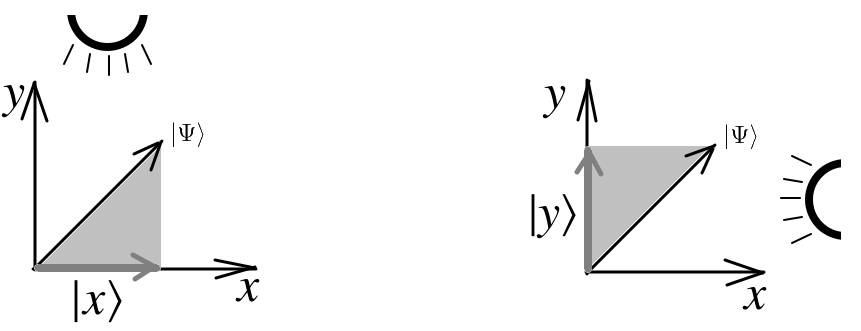} 
\captionof{figure}{The {\it shadow} of $|\Psi\rangle$ in the X-axis, $|x\rangle$, and in the Y-axis, $|y\rangle$.}
\end{center}

The concept of ``subsystem'' and that of ``product state'' lead to the set-theoretic idea that the ``system'' is the union of its parts --which, as it is very well known, is not the case in the quantum formalism and vectorial spaces where there are no `parts' (i.e., elements of a set) but `shadows' (i.e., projection of subspaces). To say that the projections of subspaces should be understood --following the classical line of reasoning-- as ``subsystems'' in SQM leads then to essential contradictions.

\section{The Atomist Myth within Quantum Philosophy}


Philosophy of QM was born as an independent field of research during the late 1970s with the appearance of specialized philosophical journals and international meetings that became popular during the mid 1980s, specially in Europe. This new field of research had of course a very strong input from the new experimental testing of quantum entanglement by John Clauser and Alain Aspect that, through Bell inequalities, went back to the famous arguments presented by Einstein and Schr\"odinger which had been mostly banned from physics. Philosophers of QM would then focus --following the influential debate by Wigner, Everett and DeWitt-- in the creation of new ``interpretations'', narratives that would describe what QM was really talking about. However, none of these interpretations would seek to replace the ground-zero, ``standard'' reference of the theory --established during the 1930s by Bohr and Dirac-- to a microscopic realm composed by electrons, protons, and the like particles. In fact, even those interpretations that would go as far as changing the mathematical formalism of the theory --such as Bohmian mechanics and GRW-- would still take for granted the orthodox presupposition that the theory of quanta described ``small quantum particles''. No philosopher of QM would dare to claim that ``what physicist say is completely wrong'', or, more explicitly, that ``the theory does not talk about electrons, neutrons or protons.'' Instead, what philosophers of QM would do is to introduce new interpretational layers that would attempt to supplement the mainstream discourse already applied within the contemporary research in quantum physics in order to better understand quantum particles. This becomes particularly explicit when considering the specific problems addressed within the philosophical literature. Maybe the most famous of them is the so called ``measurement problem of QM'', namely, the problem of making sense of ``collapses'' and the {\it projection postulate} famously added by Dirac in a completely {\it ad hoc} manner. Taking the atomist discourse as a presupposition, Dirac assumed that a ``click'', a single outcome, expresses the presence (or absence) of a particle. Thus, a unilateral focus on the explanation of the single, unique, outcome was produced (insted of focusing in the intensive patterns that were considered by Heisenberg when he developed matrix mechanics). But how to explain the transition from the quantum superpositions to those single outcomes (that appeared as the main thing to explain, since they would express the presence of a presupposed particle)? It was to somehow fill that abyss --mainly generated by the atomist supposition-- that the projection postulate was invented. Ever since, philosophers of QM have devoted enormous efforts to the resolution or understanding of this ``measurement problem'', which is in fact only a problem when particles are assumed beforehand.

Another kernel problem in the philosophical literature involves the so called ``quantum to classical limit'' which attempts --but has never been successful-- to explain the way in which quantum particles end up generating tables, chairs and dogs. It is interesting to notice that one of the very few effects of the philosophical debate on QM with respect to mainstream physics was produced during the 1990s when the models of decoherence were severely criticized within philosophical journals. However, quite regardless of the lack of any consistent response, physicists were able to turn this failure into a new solution ``FAPP'', namely, a solution ``For All Practical Purposes''. Today, mainstream physicists --and many philosophers--, regardless of the many problems and inconsistencies, take for granted the existence of this process known as decoherence. 

Last but not least, there are numerous debates which, also taking for granted the existence of quantum particles, attempt to characterize them in terms of some of their basic features, i.e., identity, separability, locality, individuality, etc. Each one of these properties of particles has produced a problem when considered in terms of the mathematical formalism of the theory. However, these negative results, instead of being understood as pointing to the necessary rejection of the atomist representation, have been reinterpreted in ``positive'' terms, as providing some of the characteristics of these quantum entities, helping thus to characterize ``quantum particles'' in purely negative or even paradoxical terms, as non-identical, non-separable, non-local, non-individual, etc.

To conclude, the atomist metaphysics that is --consciously or not-- constantly supposed by physicists as well as by philosophers of QM has functioned as a great factory of false problems and misdirections that have concentrated the attention of researchers during the last decades.

\section{The Invariant-Objective Character of Physical Individuality}

In the history of physics, the question of individuality, the notion of an individual, arouse in the context of determining, in each theory, a state of affairs that would remain the \emph{same} across different reference frames. It is this discussion that allows us to determine what the objective referents of a theory are. Apart from the Ancient Greeks who would attempt to determine \emph{moments of unity} in general metaphysical terms, in modernity Galileo, Newton and many others would take a step further in an attempt to provide a mathematical representation of {\it  the same} independently of different viewpoints. Different perspectives even though could provide different representations should consistently refer to the same {\it moment of unity} the theory talks about. In order to do so, the notion of {\it invariance} was presented as an essential element of theoretical physical representation itself. In classical physics it was the {\it Galilean transformations} which secured the possibility to discuss about {\it the same} from different {\it reference frames}, allowing to conceive consistently and systematically a state of affairs. This physical debate about invariance would be reintroduced and reframed in the 20th century with Einstein's analysis of the {\it principle of relativity}. Einstein's starting point would be the incompatibility among three different conditions if they were maintained simultaneously: between the principle of relativity (the requirement to consistently translate the experiments in one reference frame to another equivalent one), the experimental finding of the constant speed of light, and Galilean transformations.  Contrary to the ``common sense'' choice to hold tight to the Galilean transformations --and thus retain absolute Newtonian space and time--, Einstein would prefer to retain the other two: the experimental evidence about the speed of light and, above all, the principle of relativity, without which an objective physical representation would become impossible. This was, in fact, the very basic cornerstone of any physical representation, the possibility to discuss an experiment which would remain equivalent when replicated in different laboratories or reference frames. In the case of relativity theory the price to pay for sustaining the consistent translation of experience between different yet equivalent reference frames would be the {\it relativization} of the ``commonsensical'' notions of absolute space and time. To sustain the possibility and consistency of an objective physical theory, Einstein was forced to introduce a conceptual innovation, alien to classical physics. Fidelity to the irreducible conditions of a physical theory seemed rightly more important to him than fidelity to the concepts of classical theory. As a consequence, while in classical mechanics the spatial and temporal values were considered absolute (independent of reference frames) and speed and position as relative (to reference frames), in relativity theory it is the speed of light that would become absolute (independent of reference frames) and spatial and temporal intervals relative (to each reference frame). Of course, it should be stressed that the relative aspects involved in both theories did not involve any inconsistency since in each case the relative values of properties would be consistently considered in terms of a {\it global transformation}, namely, the Galilean transformation in the case of classical mechanics and the Lorentz transformation in the case of special relativity. And it is this global aspect provided by invariance that would allow in both cases to retain a consistent representation where all reference frames remain completely equivalent, this is, related to the same objective state of affairs. Even though the values of velocities and position in classical mechanics as well as temporal and spatial intervals in relativity theory are {\it relative} to the choice of a specific reference frame, the mathematical transformations secure the global consistency of the theoretical representation of {\it the same} state of affairs which, in this sense, remains completely independent of the choice of the particular referent frame. All referent frames are equivalent and can be computed from one another. 

In this respect, Bohr's approach to the theory of quanta might be considered as a diametrically opposite path. Indeed, while Einstein had given up the commonsensical image provided by classical mechanics in spatiotemporal terms in order to save the principle of relativity --and thus, retain a consistent representation between the different yet equivalent reference frames-- in the case of Bohr exactly the inverse methodology was applied in the theory of quanta. In order to save the individuals of classical physics (particles and waves) and in general to maintain the general frame of classical concepts, Bohr was willing to abandon the invariant elements of the mathematical formalism. In this respect, Bohr \cite[p. 7]{WZ} was ready to stress that: ``[...] the unambiguous interpretation of any measurement must be essentially framed in terms of classical physical theories, and we may say that in this sense the language of Newton and Maxwell will remain the language of physicists for all time'', and consequently, ``it would be a misconception to believe that the difficulties of the atomic theory may be evaded by eventually replacing the concepts of classical physics by new conceptual forms.'' Thus, Bohr, in order to hold tight to the classical notions of `wave' and `particle' (thus bringing from the outside a pre-theorical {\it moment of unity} alien to the theory, dogmatically imposed, not produced through invariance), willingly abandoned the objective-invariant consistency required by theoretical physical representations. Instead of remaining faithful to the methodological conditions that allow for an objective physical representation --a fidelity that, as in the case of Einstein, would entail to criticize some of the presuppositions inherited from the classical representation--, Bohr would choose to dogmatically hold tight to the language and the representation of classical physics, and was willing to pay the price of destroying the conditions of invariance and objectivity for the theory of quanta.  According to Bohr, QM had to be understood as a generalization of classical physics, and referred to a microscopic world composed of elementary particles that could also behave as waves \cite{Bokulich05}. According to Bohr quantum objects would require different complementary --yet incompatible-- representations such as that of `wave' and `particle' that would come to explicitly depend on the choice of the particular experimental context (or, in mathematical terms, a basis given by a complete set of commuting observables). Destroying objectivity (i.e., the categorical construction of a {\it moment of unity} capable to account consistently for different phenomena) Bohr's notion of complementarity would impose an inconsistent contextual representation where the pre-requisite to discuss about quantum objects would be the actual effectuation of the experimental arrangement. As he would famously argue in his reply to the EPR paper: 
\begin{quotation}
\noindent {\small ``it is only the mutual exclusion of any two experimental procedures, permitting the unambiguous definition of complementary physical quantities, which provides room for new physical laws, the coexistence of which might at first sight appear irreconcilable with the basic principles of science. It is just this entirely new situation as regards the description of physical phenomena, that the  notion of complementarity aims at characterizing.'' \cite[p. 700]{Bohr35}}
\end{quotation}

In turn, the destruction of objectivity would be mathematically supplemented by Dirac's re-definition of the notion of (quantum) state --also in contextual terms-- as dependent of a specific basis. In this way, the notion of {\it preferred basis} --in radical contraposition to the {\it principle of relativity}-- would become essential within mainstream quantum physics and philosophy. It is essential to recognize that  in this case, the relativism involved would differ significantly from that imposed by classical physics and --even-- relativity theory with respect to reference frames. As we argued above, while in the case of classical physics and relativity the relative values with respect to different reference frames could be consistently and invariantly considered from a global transformation allowing a common reference to the same state of affairs independent of reference frames, in the Bohr-Dirac account of QM each different reference frame would come to describe a different state of affairs incompatible with the others. Thus, while in the first case what is relative to the different reference frames is not really decisive, since we are capable of consistently considering the different frame-dependent representations, in the latter case we might talk about the introduction of a fundamental {\it perspectival relativism} in the core of the theory, where the impossibility of a global consistent account is simply established as a feature of the theory itself. Bohr would then simply replace objectivity and invariance by complementarity and contextuality. As it is well known, the formal destruction of the consistency of values considered from different reference frames would be explicitly demonstrated in the late 1960s by the famous Kochen-Specker theorem \cite{KS}. However, let us remark that Bohr's approach, which finally ended in the establishment of SQM, was not the only possible path to follow. In fact, as we will discuss in the following section, the original mathematical formalism of the theory --which had been actually developed by Heisenberg departing explicitly from Bohr's atomist program-- was designed to secure an invariant account of quantum phenomena and thus, opened the doors to an a new conceptual (non-classical) representation.

\section{Atomism or Invariance in Quantum Mechanics?}

What would it mean to take another path, to not assume the atomist representation, and to advance in a different representation, one that is also capable of respecting the basic methodological conditions for a consistent physical theory? In our opinion, the guide is given by the concept of invariance, and, more specifically, the formal invariance we find in Heisenberg's matrix mechanics. At this point it is interesting to notice that the quantum mechanical formalism was born in the year 1925 out of an explicit rejection of atomism. For some years, Heisenberg had followed Bohr’s guide, focusing on the question of describing the trajectories of electrons inside the atom --pictured by the Danish physicist as a small planetary system with discrete orbits. However, the critical reaction of Wolfgang Pauli and Arnold Sommerfeld convinced him to take a radically different path \cite{BokulichBokulich20}. So, instead of trying to describe trajectories of unseen, presupposed, corpuscles, Heisenberg reframed the problem in terms of observable quantities. As explained by Jaan Hilgevoord and Joos Uffink \cite{HilgevoordUffink01}: “His leading idea was that only those quantities that are in principle observable should play a role in the theory, and that all attempts to form a picture of what goes on inside the atom should be avoided. In atomic physics the observational data were obtained from spectroscopy and associated with atomic transitions. Thus, Heisenberg was led to consider the ‘transition quantities’ as the basic ingredients of the theory.” That same year, he would present his groundbreaking results stressing his positivist standpoint \cite{Heis25}: “In this paper an attempt will be made to obtain bases for a quantum-theoretical mechanics based exclusively on relations between quantities observable in principle.” Emancipating himself completely from the atomist picture, Heisenberg was able to create a completely new mathematical formalism. As he would recall in his autobiography:
\begin{quotation}
\noindent {\small ``In the summer term of 1925, when I resumed my research work at the University of G\"ottingen --since July 1924 I had been {\it Privatdozent} at that university-- I made a first attempt to guess what formulae would enable one to express the line intensities of the hydrogen spectrum, using more or less the same methods that had proved so fruitful in my work with Kramers in Copenhagen. This attempt led me to a dead end --I found myself in an impenetrable morass of complicated mathematical equations, with no way out. But the work helped to convince me of one thing: that one ought to ignore the problem of electron orbits inside the atom, and treat the frequencies and amplitudes associated with the line intensities as perfectly good substitutes. In any case, these magnitudes could be observed directly, and as my friend Otto had pointed out when expounding on Einstein's theory during our bicycle tour round Lake Walchensee, physicists must consider none but observable magnitudes when trying to solve the atomic puzzle.'' \cite[p. 60]{Heis71}}
\end{quotation}

Heisenberg was capable of developing matrix mechanics following two ideas: first, to leave behind the classical notion of particle-trajectory, as it did not seem required by QM --and it rather appeared as a classical habit that was inadvertently coloring a priori the approach to the understanding of the new theory--, and second, to take as a methodological standpoint Ernst Mach’s positivist idea according to which a theory should only make reference to what is actually observed in the lab, which in this case were different spectrums of line intensities. This is what was described by the tables of data that Heisenberg attempted to mathematically model and that finally led him --with the help of Max Born and Pascual Jordan-- to the development of the first consistent mathematical formulation of the theory of quanta. Let us stop to take note once again of some of the conditions that were fundamental for the development of the quantum formalism. First, Heisenberg’s abandonment of Bohr’s atomist narrative and research program which focused in the description of unobservable trajectories of presupposed yet irrepresentable quantum particles. Second, the consideration of Mach’s observability principle as a methodological standpoint that --even if Heisenberg didn't fully embraced the positivist credo-- allowed him to find a starting point unburdened of those classical presuppositions. That methodological standpoint led him finally to the replacement of Bohr’s fictional trajectories (of irrepresentable electrons) by the consideration of the intensive line spectra that were actually observed in the lab. And these quantities, once detached from a supposedly necessary reduction to atomic elements, were what the formalism was indicating as invariant. Radically new, and of fundamental importance to produce a consistent and invariant quantum formalism, was this idea that we should accept intensive values as basic, as perfectly good “substitutes”, that are in no need whatsoever to be reconducted to binary values. Intensities appeared as basic and sufficient, they seemed to require to be taken seriously, not as secondary values dependent on particles, and they seemed to point to the necessary development of a concept of an originally intensive physical element, to which the theory would primarily refer. Again: a fidelity to invariance and to the necessary conditions of a consistent physical representation seemed, as in the case of Einstein, to entail a conceptual innovation, alien to the classical theory. But Heisenberg's intuition, according to which we should take as mainly significant the intensive patterns, was immediately discarded. 

This disregard for the invariance of intensities --which would have allowed the development of the quantum theory, starting from that invariance, through a methodological path similar to that of previous theories-- happened, among other reasons, due to a binary approach that was explicitly imposed within the axiomatization of the theory presented by Dirac, and fundamentally caused by the atomist assumption. Dirac's exposition combines a use, from the outset, of the atomist representation and language, with a positivist declaration of principles, according to which the central focus in a physical theory should be only actual observations. But this positivism is from the outset contaminated by the atomist assumption, and this makes him place unilateral emphasis on single, binary outcomes, as these seem to him to evidently represent the presence (or absence) of a particle in each case. That is to say, because of the atomist assumption, he took --contrary to Heisenberg-- the single outcomes --and not the intensive patterns--, and thus a binary vision of observation, as the central focus, as that which is above all important to explain, to salvage. But if we were to start, on the contrary, not from an atomist assumption external to the theory, but from what the theory itself, its formalism, indicates as invariant, things should be very different. Given that the theory indicates intensities as invariants, if we were to verify the existence of those intensive physical elements through particular experiments in which a single outcome is obtained at a time, of course, we would need to repeat the experiment to give a precise measure of the element considered --in which case we would indeed obtain a precise quantification of the intensity. And in this case, the single, binary outcome would not be the central thing to explain, but on the contrary, a minimal information, an insufficient measure. The proof of the inadequacy of this atomist-binary view is given, of course, by the Kochen-Specker theorem, which tells us that by advancing in a binary consideration, we lose the possibility of a consistent global valuation of the state of affairs, and therefore lose the possibility of determining a referent that remains the same across different perspectives. Thus, the atomist assumption is sustained even at the cost of destroying the invariance (which was present in matrix mechanics) and the possibility of determining an objective state of affairs.

Of course, this is what Bohr didn't want to see when he tried to spread an interpretation of relativity theory that helped him to legitimize his own perspectival relativism for quantum theory. For example, in his Commo paper from 1929 he would write: 
\begin{quotation}
\noindent {\small ``While the theory of relativity reminds us of the subjective character of all physical phenomena, a character which depends essentially upon the state of motion of the observer, so does the linkage of the atomic phenomena and their observation, elucidated by the quantum theory, compel us to exercise a caution in the use of our means of expression similar to that necessity in psychological problems where we continually come upon the difficulty of demarcating the objective content.'' \cite[p. 116]{Bohr34}}
\end{quotation} 
This either deep misunderstanding or rhetoric misdirection is also present in Bohr's famous reply to the EPR paper published in 1935 where he argued that: 
\begin{quotation}
\noindent {\small ``The dependence on the reference system, in relativity theory, of all readings of scales and clocks may even be compared with the essentially uncontrollable exchange of momentum or energy between the objects of measurements and all instruments defining the space-time system of reference, which in quantum theory confronts us with the situation characterized by the notion of complementarity. In fact this new feature of natural philosophy means a radical revision of our attitude as regards physical reality, which may be paralleled with the fundamental modification of all ideas regarding the absolute character of physical phenomena brought about by the general theory of relativity.'' \cite[p. 702]{Bohr35}}
\end{quotation}
Both fragments expose the deep failure of Bohr to understand or accept the essential role of operational-invariance within physical theories as one of the main pre-conditions for a realist ``detached subject'' representation; something which of course was replaced by his own perspectival scheme. The analogy that Bohr attempted to make between relativity and his own perspectivalist account of QM is obviously wrong: relativity theory is in no way different from classical mechanics or electromagnetism when considering its operational-invariant representation. None of these theories reminds us of the ``subjective character of all physical phenomena'' nor ``compel us to exercise a caution in the use of our means of expression'' due to their ``difficulty of demarcating the objective content.'' On the very contrary, these theories are all objective and operationally invariant, allowing thus for a detached subject representation of a state of affairs. The only difference between classical mechanics and relativity is that while in the first case it is the Galilean transformations which allows us to consider all reference frames as consistently referring to {\it the same} state of affairs, in the latter case this is done through the Lorentz transformations. As Max Jammer emphasized: 
\begin{quotation}
\noindent {\small ``Bohr overlooked that the theory of relativity is also a theory of invariants and that, above all, its notion of `events,' such as the collision of two particles, denotes something absolute, entirely independent of the reference frame of the observer and hence logically prior to the assignment of metrical attributes.''  \cite[p. 132]{Jammer74}}
\end{quotation}

\section{Following the Thread of Intensive Invariance}

From what has been said so far, we can deduce a path forward that involves, on the one hand, a conceptual innovation, and, on the other hand, a fidelity to the conditions for constructing what have proven to be successful theories such as classical mechanics, electromagnetism and relativity. We are making reference here to operational invariance and objectivity. Let us note that what has been done so far regarding QM has been exactly the opposite: a fidelity to a dogmatic representation alien to the theory, and, consequently, an abandonment of these basic physical pre-conditions. It is then a matter of starting from the invariance of intensities present in the matrix formalism and developing it towards a conceptual representation, without presupposing an atomist worldview. We are then faced with the need to develop an originally intensive physical concept, one that does not need to be reduced to an atomist narrative, and that is not committed to a binary valuation --which, as we have seen, only comes from presuposing the atomist picture. Following this path, the concept of \emph{power}, \emph{intensive power}, or \emph{power of action} has been developed. Just like classical mechanics talks about particles and electromagnetism talks about electromagnetic waves, QM speaks of \emph{powers of action}, intensive quantities of action which remain invariant across different perspectives and can be experimentally verified always with complete  (intensive) certainty. The (intensive) value corresponding to each power is what we call an  \emph{intensity} or \emph{potentia}. Such values are obtained from intensive patterns which in some cases are produced through the repetition of single outcomes. Let us discuss this in formal mathematical terms. 

We begin with the definition of a (simple) \emph{graph} as a pair $G = (V, E)$, where $V$ is a set whose elements are called vertices (or nodes), and $E$ is a set of unordered pairs $\{v,w\}$ of vertices, whose elements are called edges. While each {\it vertex} is related to the mathematical notion of {\it projector operator} and to the physical concept of  {\it power of action}, each {\it edge} is linked to the mathematical concept of {\it commutation} and the physical compatibility of powers within an experimental arrangement. 

\begin{definition} {\bf Graph of powers:} Given a Hilbert space $H$, the graph of powers $G(H)$ is defined such that the vertices are the projectors on $H$ (called \emph{powers}), and an edge exists between projectors $P_1$ and $P_2$ if they commute.
\end{definition} 

\noindent It is these powers, in their multiplicity and their relationships, that allow us to define an \emph{Intensive State of Affairs} (ISA) --in contrast to an Actual (Binary) State of Affairs (ASA). But first, we need to formalize the notion of intensity (or potentia). The assignment of intensities is called Global Intensive Valuation (GIV).

\begin{definition} {\bf Global Intensive Valuation:}  A Global Intensive Valuation is a map from $G(H)$ to the interval $[0,1]$.
\end{definition}

\noindent Clearly, not all GIVs are compatible or consistent with the relations between powers. We will focus on those that define an ISA as follows:

\begin{definition} {\bf Intensive State of Affairs:} Let $H$ be a Hilbert space of infinite dimension. An \emph{Intensive State of Affairs} is a GIV $\Psi: G(H)\to[0,1]$ from the graph of powers $G(H)$
such that $\Psi(I)=1$ and 
\[
\Psi(\sum_{i=1}^{\infty} P_i)=\sum_{i=1}^\infty \Psi(P_i)
\]
for any piecewise orthogonal operator $\{P_i\}_{i=1}^{\infty}$. The numbers $\Psi(P) \in [0,1]$ are called {\it intensities} or {\it potentia} and the vertices $P$ are called \emph{powers of action}. Taking into consideration the ISAs, it is then possible to advance towards a consistent GIV which can bypass the contextuality expressed by the Kochen-Specker Theorem \cite{deRondeMassri21a, KS}. 
\end{definition} 
At this point, it is necessary to say that a fundamental characteristic of any conceptual representation of a physical theory is its operationality. That is, the ability to relate physical concepts to what is actually observed in experiments. As Einstein explained, a physical concept lacks value if we are unable to connect it with an experimental corroboration. 
\begin{definition} {\bf Quantum Laboratory:} We use the term \emph{quantum laboratory} (or quantum lab or Q-Lab) as the operational concept of an ISA. 
\end{definition} 
\begin{definition}{\bf Screen and Detector:} A \emph{screen} with $n$ places for $n$ detectors corresponds to the vector space $\mathbb{C}^n$. Choosing a basis, say $\{|1\rangle,\dots,|n\rangle\}$, is the same as choosing a specific set of $n$ {\it detectors}. A \emph{factorization} $\mathbb{C}^{i_1}\otimes\dots \otimes\mathbb{C}^{i_n}$ is the specific number $n$ of screens, where the screen number $k$ has $i_k$ places for detectors, $k=1,\dots,n$. Choosing a \emph{basis} in each factor corresponds to choosing the specific detectors; for instance $|\uparrow\rangle, |\downarrow\rangle$. After choosing  a basis in each factor, we get a basis of the factorization $\mathbb{C}^{i_1}\otimes\dots \otimes\mathbb{C}^{i_n}$
that we denote as
\[
\{ |k_1\dots k_n\rangle \}_{1\le k_j\le i_j}.
\]
\end{definition} 

\begin{definition}{\bf Power of action:} The \emph{ basis element} $|k_1\dots k_n\rangle$ determines the \emph{ projector}  $|k_1\dots k_n\rangle \langle k_1\dots k_n|$ which is the formal-invariant counterpart of the objective physical concept called \emph{ power of action} (or simply \emph{power}) that produces a global effect in the $k_1$ detector of the screen $1$,  in the $k_2$ detector of the screen $2$ and so on until the $k_n$ detector of the screen $n$. Let us stress the fact that this effectuation does not allow an explanation in terms of particles within classical space and time. Instead, this is explained as a characteristic feature of powers. In general, any given power will produce an intensive multi-screen non-local effect. 
\end{definition} 

\begin{definition} {\bf Experimental Arrangement:} Given an ISA, $\Psi$, a factorization $\mathbb{C}^{i_1}\otimes\dots \otimes\mathbb{C}^{i_n}$ and a basis $B=\{|k_1\dots k_n\rangle\}$ of cardinality $N=i_1\dots i_n$, we define an \emph{ experimental arrangement} denoted $\EA_{\Psi,B}^{N,i_1\dots i_n}$, as a specific choice of screens with detectors together with the potentia of each power, that is,
\[
\EA_{\Psi,B}^{N,i_1\dots i_n}= \sum_{k_1,k_1'=1}^{i_1}\dots \sum_{k_n,k_n'=1}^{i_n} 
\alpha_{k_1,\dots,k_n}^{k_1',\dots,k_n'}|k_1\dots k_n\rangle\langle k_1'\dots k_n'|.
\]
Where the number $N$ is the cardinal of $B$ and is called the \emph{degree of complexity} (or simply \emph{degree}) of the experimental arrangement. 
\end{definition} 
\begin{definition}{\bf Potentia:} The number that accompanies the power $|k_1\dots k_n\rangle \langle k_1\dots k_n|$ is its \emph{ potentia} (or intensity) and the basis $B$ determines the powers defined by the specific choice of screens and detectors. 
\end{definition} 

In this way, starting solely from the operational-invariance present in the quantum formalism it is possible to develop --following the general conditions discussed above-- a representation that is not indebted to the atomist picture and that allows to escape contextuality --namely, the {\it perspectival relativism} orthodoxly considered as an essential feature of QM-- as well as the need for ``collapses''. We can thus say that, methodologically, QM is not very different from the theories of Newton, Maxwell or Einstein, but it is of course conceptually different --something which is neither new. What is truly different is the way in which QM forces us to consider the individuals of the theory in relative terms.

\section{The Relative Nature of Quantum Individuals}

Given our analysis, just in the same way that Einstein was confronted with a choice between the principle of relativity, the experimental finding of the invariance  of the speed of light and the absolute nature of space and time --inherited from the classical representation--, when considering QM we are also faced with an incompatibility between the intensive nature of quantum phenomena, the invariant mathematical formalism and the atomist representation, with its reference to particles and all its accompanying temptations --also inherited from the classical representation. In both cases something has to go in order to retain the consistency of the theory. But while Einstein would choose to abandon --in order to come up with special relativity-- our ``common sense'' spatiotemporal picture and retain the experimental evidence of the speed of light as well as the equivalence between reference frames, Bohr would prefer in the theory of quanta to retain the atomist understanding of individuality and give up instead not only the intensive phenomena observed in the lab (replaced by binary values that expressed the observation of ``corpuscles'') but also the invariant consistency of the mathematical formalism itself (destroyed through the reference to binary values and the introduction of ``collapses'').  Today, after almost a century of the establishment of Bohr's orthodox account of QM we might critically reconsider the results of having followed this latter path. In fact, Bohr's account of QM has failed to produce a deeper understanding of the theory and has lead us instead to a dangerous situation where fragmentation, vagueness and inconsistency are threatening the contemporary research not only within theoretical developments but also within technological ones. Thus, it might be time to propose something different, a different path that might be grounded on the successful approaches of the past. It is in this context that we propose here to follow the same approach which lead Einstein to the successful development of relativity theory --as Newton and Maxwell had done before-- and hold fast to the general theoretical pre-conditions that constrain physical representation itself, namely, consistent operational invariance and experience while at the same being prepared to advance in a conceptual scheme which departs from our classical account of reality. 

Following Einstein's guidance in this case means setting aside the classical, atomistic conception of individuality. As shown previously, we start from the need to develop the concept of an originally intensive physical element. In this sense, the concept of power (or intensive power, or power of action) was proposed, consistently accompanied by the concept of intensity (the intensive quantification in each case of a power), by an Intensive State of Affairs, and thus a consistent Global Valuation that precludes contextuality right from the start. Now, when we think about how to define individuality in this context, we encounter profound differences with the classical scheme, differences which --in turn-- force us to conceive new ways of thinking. Classical physics (as well as the Bohrian path in QM) implies an absolutization of the individual, in the sense that the individual (the particle for example) is given as an independent, separate substantial entity that completely constitutes the totality in its mere sum. Fundamental reality is given by separate, independent, substantial individual entities. On the contrary, in QM, when we determine what remains the same, we encounter intensive physical elements that can no longer be understood as substantial separated entities, as in the atomist representation. An ISA cannot be defined (for both conceptual and formal reasons) as a mere sum of independent and separate entities. There is no separate, independent, isolated power of action. Powers always come in interrelated multiplicities. A power cannot be defined in the atomistic manner. We thus encounter a fundamentally and naturally relational, systemic scheme, an originally relational conception. There is a co-belonging, a mutual dependence of all powers that is fundamental, irreducible, and which defines each and every situation. Thus, just like Einstein was forced to relativize the notions of space and time in order to retain the consistency of experimental findings and mathematical transformations, we are led in the theory of quanta to relativize the notion of individual. We will never have an absolute, ontologically separated, isolated quantum individual (the only absolute individual in this scheme would be, in a Spinozist manner, the totality of nature), instead what we obtain are {\it relative individuals}. However, it is important to stress that this relative character of quantum individuals does not imply a perspectival dependence that --like in the Bohrian scheme-- precludes a global representation which consistently unites what is obtained form different reference frames. This relative character of quantum individuals is objective and not perspectival, it does not imply their dependence on a perspective, but rather their intrinsic relationality, their non-completeness, their non-absolute, non-separated, non-fundamental nature. Thus, a quantum individual will always be relative (non-complete, non-absolute) and will inevitably refer to a specific invariant multiplicity of interrelated powers of action. Of course, between powers and between states of affairs there are real differences, but these differences do not imply absolute ontological separations. In fact, difference is a form of relation. Now, in the context of this relational scheme, the question rises: what specific multiplicity of powers should we take as defining a relative individual? What multiplicity could be significant? Which multiplicity would make this notion of a relative individual physically and operationally useful?  

Before answering these questions we need to address two important theorems (derived in \cite{deRondeMassri23}) that allow us to consider in formal terms the relations between powers, intensities, experimental arrangements and quantum labs. Assume that in a Q-Lab we want to change or modify an experimental setup by changing the number of screens and detectors. There are two theorems that allow us to relate the different reference frames and factorizations. If the number of powers (i.e., the degree of complexity) remains the same after the rearrangement, then the {\it Basis Invariance Theorem} tell us that the new experimental arrangement is equivalent to the previous one, but if the complexity of the new experimental arrangement drops, then the {\it Factorization Invariance Theorem} tell us that all the knowledge in the new experimental arrangement was already contained in the previous one (see for a detailed analysis \cite{deRondeFMMassri24a, deRondeFMMassri24b}). 

\begin{theo}{\sc (Basis Invariance Theorem)}
Given a specific Q-Lab $\Psi$, all experimental arrangements of the same complexity, $EA_{\Psi}^N$, are equivalent independently of the basis. 
\end{theo}

\begin{theo} {\sc (Factorization Invariance Theorem)}
The experiments performed within a $EA_{\Psi}^N$ can also be performed with an experimental arrangement of higher complexity N+M, $EA_{\Psi}^{N+M}$, that can be produced within the same Q-Lab $\Psi$.  
\end{theo}
\noindent Following the {\it Factorization Invariance Theorem}, we can thus say that a more complex experimental arrangement (one which considers more powers) includes less complex experimental arrangements. Meaning that all EAs can be completely deduced from any EA that is more complex. This also implies that the more intensive powers we consider (that is, the more complex the individual), the more knowledge we have of the state of affairs. This is completely different from the orthodox case in which a pure state provides maximal knowledge. In our case, the knowledge is directly related to the complexity and not to purity. In fact, a pure state of degree 1 will provide the minimal possible information about that state of affairs.

Given these theorems we are ready to propose the following definition of {\it quantum individual}: the specific multiplicity of powers that allows, in each case, to completely determine not only a physical situation but also all its possible transformations. Or, to put it more precisely: the minimum set of powers of action within a specific degree of complexity capable of deriving the totality of powers and potentia in that same degree (or less). 

\begin{definition}{\bf  Quantum Individual:}  A quantum individual is a set of powers of complexity $N$ capable to derive the totality of powers and their respective potentia in that specific degree (or less).
\end{definition}
Something characteristic of the relative individual is that, having an EA of a certain complexity, that is, a specific set of powers, I can deduce any other EA of that same degree of complexity (or less), determined by another set of powers. Thus, having a consistent set of powers with their potentia, I can deduce all the other powers (and their respective potentia). 

Let us emphasize once again that the important thing to understand is that this representation speaks of the systematic belonging, the reciprocal relationship, of intensive physical elements, of intensive quantities of action, that is, of powers of action. Beyond the terminology used (whose consistency is however always necessary), the central point to understand is the basic intensive nature of the physical elements, that is, their non-dependence on a supposedly more fundamental atomistic representation. The power of action is not an action of a substantial body. In analogous terms to electromagnetism, there are no subjects nor objects on which powers depend or need to be attached to. That is their original character. It is also central to assume the originally relational or systemic nature of the representation, where the basic, foundational role is not held by absolute, independent individuals that would constitute the totality as a sum. There is a systemic or relational totality that is simply irreducible. These relative individuals are, of course, not arbitrary, in the sense, first, that it is necessary to find the significant relational multiplicities, and, second, that they are invariant, meaning that their definition and valuation are consistently related through the different reference frames, globally. In this representation, a specific power, with its determined intensive value, relates to other specific powers not in just any way but according to the determined form (the relational system) of a particular state of affairs to which it inevitably belongs. That is, not everything relates to everything else in any way; there are determined relationships between powers. This is how real difference (the determined and valuable reality of each power) and irreducible relation (the fact that powers always come in relations) are articulated, thus defining not only specific and precisely verifiable situations but also a unified consistent state of affairs. In this way, we can also say that a determined power of action which is part of a particular state of affairs does not relate on its own to another determined power of a different state of affairs. They can only do so, more indirectly, through the consideration of the relationship between those two states of affairs as a whole (that is, through the consideration of a larger relative individual that includes the relative individuals in which those powers are found).

To conclude, let us examine a specific example showing the application of the theorems and the physical concepts. Let us work within Q-Lab $\Psi$ with two screens and two detectors. Formally, this is saying that we are working with the factorization $\mathbb{C}^2\otimes\mathbb{C}^2$. Let us choose some specific detectors in each screen, $B_1 = \{|1\rangle,|2\rangle\}$ and $B_2 = \{|1\rangle,|2\rangle\}$. Then, this experimental arrangement has four possible combinations of detectors defining the basis 
$B =  \{|11\rangle,|12\rangle, |21\rangle,|22\rangle\}$ and the powers
\[
|11\rangle\langle 11|,|12\rangle\langle 12|, |21\rangle\langle 21|,|22\rangle\langle 22|.
\]
Les us assume that in the chosen basis $B$, the experimental arrangement is given by
\[
\EA_{\Psi,B} = \frac{2}{10}|11\rangle\langle 11| +\frac{3}{10}|12\rangle\langle 12| +\frac{1}{10}|21\rangle\langle 21|+\frac{4}{10}|22\rangle\langle 22|
\]
only with non zero diagonal terms. Now, let us compute an equivalent experimental arrangement (the equivalence follows from the {\it Basis Invariance Theorem}) in a new basis $B'$ by changing the detectors in the second screen in the following manner: $|1\rangle = (|+\rangle + |-\rangle)/\sqrt{2}$ and $|2\rangle = (|+\rangle - |-\rangle)/\sqrt{2}$. We can then compute the different relations between the powers, 
\begin{align*}
|11\rangle\langle 11| &=
\frac{1}{2}|1+\rangle\langle 1+| + 
\frac{1}{2}|1+\rangle\langle 1-|  +
\frac{1}{2}|1-\rangle\langle 1+|  +
\frac{1}{2}|1-\rangle\langle 1-| \\
|12\rangle\langle 12| &=
\frac{1}{2}|1+\rangle\langle 1+| - 
\frac{1}{2}|1+\rangle\langle 1-|  -
\frac{1}{2}|1-\rangle\langle 1+|  +
\frac{1}{2}|1-\rangle\langle 1-| \\
|21\rangle\langle 21| &=
\frac{1}{2}|2+\rangle\langle 2+| + 
\frac{1}{2}|2+\rangle\langle 2-|  +
\frac{1}{2}|2-\rangle\langle 2+|  +
\frac{1}{2}|2-\rangle\langle 2-| \\
|22\rangle\langle 22| &=
\frac{1}{2}|2+\rangle\langle 2+| - 
\frac{1}{2}|2+\rangle\langle 2-|  -
\frac{1}{2}|2-\rangle\langle 2+|  +
\frac{1}{2}|2-\rangle\langle 2-| 
\end{align*}
Then, 
\begin{align*}
\EA_{\Psi,B'}& =  
\frac{1}{4}|1+\rangle\langle 1+| 
-\frac{1}{20}|1+\rangle\langle 1-|  
-\frac{1}{20}|1-\rangle\langle 1+|  
+\frac{1}{4}|1-\rangle\langle 1-|\\
&+\frac{1}{4}|2+\rangle\langle 2+| 
-\frac{3}{20}|2+\rangle\langle 2-| 
-\frac{3}{20}|2-\rangle\langle 2+| 
+\frac{1}{4}|2-\rangle\langle 2-|.
\end{align*}
Notice that the intensities of the powers in this new experimental arrangement are $(1/4,1/4,1/4,1/4)$, but on the previous one, the intensities were $(2/10,3/10,1/10,4/10)$. Here, due to the {\it Basis Invariance Theorem} we are certain that both EA are equivalent. 

Now, let us consider the first screen as an experimental arrangement in itself, where the detectors are the same as before $B_1 = \{|1\rangle,|2\rangle\}$. We can compute the new intensities in two equivalent ways (the equivalence follows from the {\it Factorization Invariance Theorem}) either using the first experimental arrangement $\EA_{\Psi,B}$ or the second one $\EA_{\Psi,B'}$. By using the partial trace of the first factor $\text{Tr}_2:B(\mathbb{C}^2\otimes\mathbb{C}^2)\to B(\mathbb{C}^2)$ we have the following relations between powers,
\begin{align*}
\text{Tr}_2(|11\rangle\langle 11|)=
\text{Tr}_2(|12\rangle\langle 12|)=
\text{Tr}_2(|1+\rangle\langle 1+|)=
\text{Tr}_2(|1-\rangle\langle 1-|)&=|1\rangle\langle 1|,\\
\text{Tr}_2(|12\rangle\langle 11|)=
\text{Tr}_2(|11\rangle\langle 12|)=
\text{Tr}_2(|1-\rangle\langle 1+|)=
\text{Tr}_2(|1+\rangle\langle 1-|)&=0,\\
\text{Tr}_2(|21\rangle\langle 21|)=
\text{Tr}_2(|22\rangle\langle 22|)=
\text{Tr}_2(|2+\rangle\langle 2+|)=
\text{Tr}_2(|2-\rangle\langle 2-|)&=|2\rangle\langle 2|,\\
\text{Tr}_2(|22\rangle\langle 21|)=
\text{Tr}_2(|21\rangle\langle 22|)=
\text{Tr}_2(|2-\rangle\langle 2+|)=
\text{Tr}_2(|2+\rangle\langle 2-|)&=0.
\end{align*}
Then, 
\begin{align*}
\text{Tr}_2(\EA_{\Psi,B})=&\left(\frac{2}{10}+\frac{3}{10}\right)|1\rangle\langle 1| + 
\left(\frac{1}{10}+\frac{4}{10}\right)|2\rangle\langle 2|,\\
\text{Tr}_2(\EA_{\Psi,B'})=&\left(\frac{1}{4}+\frac{1}{4}\right)|1\rangle\langle 1| + 
\left(\frac{1}{4}+\frac{1}{4}\right)|2\rangle\langle 2|.
\end{align*}
Hence, as stated before in the {\it Factorization Invariance Theorem}, both experimental arrangement $\EA_{\Psi,B}$ and $\EA_{\Psi,B'}$ provide the complete knowledge of an experimental arrangement $\EA_{\Psi,B_1}$ of lower degree.



\section*{Acknowledgements} 

This work was partially supported by the following grant: ANID-FONDECYT, project number: 3240436. The authors state that there is no conflict of interest.


\begin{thebibliography}{1}




\bibitem{Arenhart17}  Arenhart, J.R.B., 2017, ``The received view on quantum non- individuality: formal and metaphysical analysis'', {\it Synthese}, {\bf 194}, 1323-1347.

\bibitem{Arenhart23} Arenhart, J.R.B., 2023, ``A no-individuals account of quantum mechanics'', {\it Philosophical Transactions A}, {\bf 381}, 20220097.

\bibitem{Bacc12} Bacciagaluppi, G., 2012, ``The Role of Decoherence in Quantum Mechanics'', {\it The Stanford Encyclopedia of Philosophy (Winter 2012 Edition)}, Edward N. Zalta (ed.), http://plato.stanford.edu/archives/win2012/entries/qm-decoherence/.

\bibitem{BvN36} Birkhoff, G. $\&$ von Neumann, J., 1936, ``The logic of quantum mechanics'', {\it Annals of Mathematics}, {\bf 37}, 823-843.

\bibitem{Bohr34} Bohr, N., 1934, {\it Atomic Theory and the Description of Nature}, Cambridge University Press, Cambridge.

\bibitem{Bohr35} Bohr, N., 1935, ``Can Quantum Mechanical Description of Physical Reality be Considered Complete?'', {\it Physical Review}, {\bf 48}, 696-702.

 \bibitem{Bokulich05} Bokulich, A., 2005, ``Niels Bohr's Generalization of Classical Mechanics'', {\it Foundations of Physics}, {\bf 35}, 347-371.
 
\bibitem{BokulichBokulich20} Bokulich, A. $\&$ Bokulich, P., 2020, ``Bohr's Correspondence Principle'', {\it The Stanford Encyclopedia of Philosophy (Fall 2020 Edition)}, E.N. Zalta (ed.), forthcoming. https://plato.stanford.edu/archives/fall2020/entries/bohr-correspondence/.

\bibitem{BokulichJaeger10} Bokulich, A. $\&$ Jaeger, G., 2010, {\it Philosophy of Quantum Information and Entanglement}, Cambridge University Press, Cambridge. 

\bibitem{Bub97} Bub, J., 1997, {\it Interpreting the Quantum World}, Cambridge University Press, Cambridge.

\bibitem{Nature22} Castelvecchi, D.  $\&$ Gibney, E., 2022, ```Spooky' quantum-entanglement experiments win physics Nobel'', {\it Nature}, {\bf 610}, 241-242.

\bibitem{CleveBuhrman97} Cleve, R.  $\&$ Buhrman, H., 1997, ``Substituting quantum entanglement for communication'', {\it Physical Review A}, {\bf 56}, 1201.

\bibitem{deRondeFM21} de Ronde, C. $\&$ Fern\'andez-Mouj\'an, R., 2021, ``Are `Particles' in QM `Just a Way of Talking'?'', preprint (http://philsci-archive.pitt.edu/19968). 

\bibitem{deRondeFMMassri24a} de Ronde, C., Fern\'andez Mouj\'an, R. $\&$ Massri, C., 2024, ``Equivalence Relations in Quantum Theory: An Objective Account of Bases and Factorizations'', sent (quant-ph:2404.14891).

\bibitem{deRondeFMMassri24b} de Ronde, C., Fern\'andez Mouj\'an, R. $\&$ Massri, C., 2024, ``Everything is Entangled in Quantum Mechanics: Are the Orthodox Measures Physically Meaningful?'', sent (quant-ph:2405.05756).

\bibitem{deRondeFreytesDomenech18} de Ronde, C., Domenech, G. $\&$ Freytes, H., 2018, ``Quantum Logic in Historical and Philosophical Perspective'', {\it Internet Encyclopedia of Philosophy}, URL: https://iep.utm.edu/qu-logic/

\bibitem{deRondeMassri21a} de Ronde, C. $\&$ Massri, C., 2021, ``The Logos Categorical Approach to Quantum Mechanics: I. Kochen-Specker Contextuality and Global Intensive Valuations.'', {\it International Journal of Theoretical Physics}, {\bf 60}, 429-456. 

\bibitem{deRondeMassri23} de Ronde, C. $\&$ Massri, C., 2023, ``Relational quantum entanglement beyond non-separable and contextual relativism'', {\it Studies in History and Philosophy of Science}, {\bf 97}, 68-78.

\bibitem{Earman15} Earman, J., 2015, ``Some Puzzles and Unresolved Issues About Quantum Entanglement'', {\it Erkenntnis}, {\bf 80}, 303-337.

\bibitem{Feynman67} Feynman, R.P., 1967, {\it The Character of Physical Law}, Massachusetts Institute of Technology Press, Massachusetts. 

\bibitem{Heis25} Heisenberg, W., 1925, ``\"{U}ber Quantentheoretische Umdeutung Kinematischer und Mechanischer Beziehungen'', {\it Zeitschrift f\"{u}r
Physik}, \textbf{33}, 879-893. English translation in \emph{Sources of Quantum Mechanics}, B.L. van der Waerden, ed., Dover, 1968.

\bibitem{Heis71} Heisenberg, W., 1971, {\it Physics and Beyond}, Harper \& Row, New York.

\bibitem{HilgevoordUffink01} Hilgevoord, J. $\&$ Uffink, J., 2001, ``The Uncertainty Principle'', {\it The Stanford Encyclopedia of Philosophy (Winter 2001 Edition)}, E. N. Zalta (Ed.), URL: http://plato.stanford.edu/archives/win2001/entries/qt-uncertainty/.

\bibitem{Horodecki09} Horodecki, R., Horodecki, P., Horodecki, M. $\&$ Horodecki, K., 2009, ``Quantum entanglement'', {\it Reviews of Modern Physics}, {\bf 81}, 865-942.

\bibitem{Jammer74} Jammer, M., 1974,  {\it The Philosophy of Quantum Mechanics},  John Wiley and sons, New York.

\bibitem{KS} Kochen, S. $\&$ Specker, E., 1967, ``On the problem of Hidden Variables in Quantum Mechanics'', {\it Journal of Mathematics and Mechanics}, {\bf 17}, 59-87. Reprinted in Hooker, 1975, 293-328.

\bibitem{KrauseArenhart18} Krause, $\&$ Arenhart, J.R.B., 2018, ``Quantum Non-individuality: Background Concepts and Possibilities'', in {\it  The Map and the Territory}, Wuppuluri, S., Doria, F. (Eds) ,The Frontiers Collection, Springer, Berlin.

\bibitem{Laudan81} Laudan, L., 1981, ``Ernst Mach's Opposition to Atomism'', in {\it Science and Hypothesis}, Springer, Dordrecht.

\bibitem{Maudlin19} Maudlin, T., 2019, {\it Philosophy of Physics. Quantum Theory}, Princeton University Press, Princeton.

\bibitem{Moore89} Moore, W., 1989, {\it Schr\"odinger}, Cambridge University Press, New York.

\bibitem{NeyAlbert13} Ney, A. $\&$ Albert, D. (Eds.), 2013, {\it The Wave Function: Essays on the Metaphysics of QM}, Oxford University Press, Oxford.  

\bibitem{Zeilinger98}  Pan, J.-W., Bouwmeester, D., Weinfurter, D. $\&$  Zeilinger, A., 1998, ``Experimental Entanglement Swapping: Entangling Photons That Never Interacted'', {\it Physical Review Letters}, {\bf 80}, 3891.

\bibitem{Paz01} Paz, J.P., 2001, ``Protecting the quantum world'', {\it Nature}, {\bf 412}, 869-870.

\bibitem{Schr50} Schr\"odinger, E., 1950, ``What is an elementary particle?'', {\it Endeavor}, {\bf 9}, 109-116.

\bibitem{Shimony95} Shimony, A., 1995, ``Degree of entanglement'', {\it Annals New York Academy of Science}, {\bf 755}, 675-679.

\bibitem{Thomas21}  Thomas, R.A., Parniak, M., Ostfeldt, C.,  Moller, C.B., Barentsen, C., Tsaturyan, Y.,  Schliesser, A., Appel, J., Zeuthen, E. $\&$  Polzik, E.S., 2021, ``Entanglement between distant macroscopic mechanical and spin systems'', {\it Nature Physics}, {\bf 17}, 228-233.

\bibitem{Vedral14} Vlatko Vedral, 2014, ``Quantum entanglement'', {\it Nature Physics}, {\bf 10}, 256-258.

\bibitem{WZ} Wheeler, J. $\&$ Zurek, W. (Eds.) 1983, {\it Theory and Measurement}, Princeton University Press, Princeton.

\bibitem{Wolchover20} Wolchover, N., 2020, ``What Is a Particle?'', {\it Quanta Magazine}, URL: https://www.quantamagazine.org/what-is-a-particle-20201112/

\bibitem{Wootters98} Wootters W. K., 1998, ``Quantum entanglement as a quantifiable resource'', {\it Philosophical Transactions of the Royal Society A}, {\bf 356}, 1717-1731.


\end{thebibliography}
\end{document}